\documentclass[aps,prd,groupedaddress,amsmath,amssymb,amsfonts,nofootinbib,preprint]{revtex4-1}

\usepackage{amssymb,amsfonts}
\usepackage{mathrsfs}
\usepackage{tikz}
\usetikzlibrary{external}
\usepackage{amsthm}

\usepackage{accents}
\usepackage{graphicx}
\usepackage[utf8x]{inputenc}
\usepackage{float}
\usepackage{amsmath}
\DeclareMathAlphabet{\mathpzc}{OT1}{pzc}{m}{it}
\usetikzlibrary{patterns}
\usepackage{titlesec}
\usepackage{cases}
\usepackage{mathtools}
\usepackage{color}
\usepackage[colorlinks, allcolors=blue!70!black, linktocpage]{hyperref}
\usepackage{cleveref}
\crefname{lem}{lemma}{lemmas}
\crefname{thm}{theorem}{theorems}
\crefname{cor}{corollary}{corollaries}
\crefname{rem}{remark}{remarks}
\crefname{prop}{proposition}{propositions}

\def\ben{\begin{equation}}
\def\een{\end{equation}}
\def\bena{\begin{eqnarray}}
\def\eena{\end{eqnarray}}

\begin{document}
\bibliographystyle{plain}

\title{The Particle and Energy Cost of Entanglement of Hawking Radiation with the Final Vacuum State}

\author{Robert M. Wald}
\email{rmwa@uchicago.edu}
\affiliation{Enrico Fermi Institute and Department of Physics \\
  The University of Chicago \\
  5640 South Ellis Avenue, Chicago, Illinois 60637, USA.}

\begin{abstract}

A semiclassical analysis shows that in the process of black hole formation and evaporation, an initial pure state will evolve to a mixed state, i.e., information will be lost. One way of avoiding this conclusion without invoking drastic modifications of the local laws of physics in a low curvature regime would be for the information to be restored at the very end of the evaporation process. It is normally envisioned that this would require a final burst of particles entangled with the early time Hawking radiation. This would imply the emission of an extremely large number of particles from an object of Planck size and mass and would appear to be blatantly ruled out by energy considerations. However, Hotta, Schutzhold, and Unruh have analyzed a $(1+1)$-dimensional moving mirror analog of the Hawking process and have found that, in this model, information is restored via entanglement of the early time Hawking radiation with vacuum fluctuations in the spacetime region to the future of the event where the mirror returns to inertial motion. We analyze their model here and give a precise formulation of this entanglement by introducing the notion of ``Milne particles.'' We then analyze the inertial particle and energy cost of such an entanglement of Hawking radiation with vacuum fluctuations. We show that, in fact, the entanglement of early time Hawking radiation with vacuum fluctuations requires the emission of at least as many late time inertial particles as Hawking particles. Although the energy cost can be made small in the $(1+1)$-dimensional mirror system, this should not be the case for the $(3+1)$-dimensional evaporating black hole system. Thus, vacuum entanglement has the same difficulties as the more usual burst scenarios for attempting to avoid information loss. 

\end{abstract}
\maketitle
\tableofcontents
\section{Introduction}

The complete gravitational collapse of a body in classical general relativity is believed to always result in the formation of a black hole. As discovered by Hawking \cite{hawk}, particle creation will occur in the vicinity of the black hole and result in a thermal flux of particles to infinity. In a semiclassical analysis, this particle flux is described by an exactly thermal density matrix due to the entanglement of the Hawking particles reaching infinity with ``particles'' inside the black hole \cite{wald1}. The fact that the Hawking radiation is in a mixed state before the black hole has evaporated is not generally considered to be problematic, since the full quantum state (including the degrees of freedom inside the black hole) is pure. However, the loss of energy of the black hole due to Hawking 
radiation should result in its complete evaporation within a finite time. In a semiclassical analysis, all that remains after black hole evaporation is the mixed state of the Hawking radiation. Thus, semiclassically, an initial pure state will evolve to a mixed state. For reasons I have not been able to understand during the course of the past 40 years, this is widely viewed as being highly problematic. The conflict between this view and the semiclassical analysis is referred to as the {\it
black hole information loss paradox}. 

If in order to avoid the conclusion of information loss, the semiclassical picture that gives rise to this conclusion must be modified. As reviewed in \cite{uwrev}, there are four basic logical possibilities for doing so: (I) No black hole actually forms in collapse, e.g., there is tunneling to a ``fuzzball'' \cite{math}. (II) Major departures from semiclassical theory occur during the evaporation process, e.g., the event horizon is converted to a ``firewall'' \cite{amps}. (III) The semiclassical analysis remains valid until the black hole reaches the Planck scale, but then the black hole stops evaporating, leaving behind a ``remnant'' that keeps the total state pure. (IV) The semiclassical analysis remains valid until the black hole reaches the Planck scale, and all of the ``information'' stored within the black hole then comes out in a ``final burst.''

Possibilities (I) and (II) require general relativity and/or quantum field theory to fail catastrophically in a low curvature regime, where, {\it a priori}, one would expect these descriptions to be valid. If the ``remnants'' of possibility (III) do not interact with the outside world, it is not clear what good is done by having ``information'' present that is inaccessible; if they do, then they should be thermodynamically favored over all other types of matter and presumably should be spontaneously produced at a high rate.
Finally, if the ``final burst'' of possibility (IV) consists of emission of particles that are entangled in an ordinary way with the Hawking radiation particles, then an object of Planck mass and Planck size would have to emit as many particles---each, presumably, of Planck frequency---in the burst as there were particles in the Hawking radiation. This is not energetically possible.

However, several years ago, an analysis by Hotta, Schutzhold, and
Unruh \cite{hsu} suggested that there may be a different way of implementing possibility (IV). These authors
considered the model of a mirror in $(1+1)$-dimensions that starts in inertial motion and then
accelerates in such a manner as to emit exactly thermal Hawking-like radiation for a long period of time.
The mirror then becomes inertial again. This is an entirely ``unitary'' process---no black hole or singularity is present---
so the full state of the quantum field must be pure at all times.
The purity of the state during the emission process while the mirror is accelerating can be understood
as a consequence of entanglement of the Hawking radiation with ``partner
particles'' that are present outside of the mirror. These partner particles eventually enter the region to
the future of the event at which the mirror becomes inertial again; they then
``bounce off''
the mirror and go back out to infinity at late times. However, the state of the quantum field in the region to the future
of the event at which the mirror becomes inertial is indistinguishable from the ordinary vacuum state for that
inertial mirror. In a sense that will be made more precise in this paper, these 
``partner particles'' are, in fact, merely
vacuum fluctuations in this region. Thus, in this model, the Hawking radiation is ``purified'' by its entanglement
with vacuum fluctuations in the future region! There is no obvious energy cost to doing this. In the review \cite{uwrev},
we considered such vacuum entanglement to be a ``potentially viable'' way of avoiding information loss in black hole evaporation.

The main purpose of this paper is to analyze the inertial particle and energy cost of the sort of vacuum entanglement 
with Hawking radiation that occurs in the model of Hotta, Schutzhold, and
Unruh \cite{hsu}. Although there is no ``direct cost'' of this entanglement in that all of the ``information'' about the Hawking radiation is stored in late time vacuum fluctuations, there is an ``indirect cost'' in that these vacuum fluctuations can no longer be correlated with other vacuum fluctuations that they would have been correlated with in the global vacuum state. We will show that this implies that the number of inertial particles emitted at late times must be at least as large as
the number of Hawking radiation particles emitted at early times. In the $(1+1)$-dimensional mirror model, the energy of these late time particles can be made very small by ``turning off'' only the acceleration of the mirror (i.e., keeping its velocity unchanged rather than returning it to rest) in the final inertial era. However, if a similar vacuum entanglement occurs in the $(3+1)$-dimensional evaporating black hole case, causality requires that these particles emerge from a Planck scale region near the evaporation event. It follows that for evaporating black holes, the particle and energy cost of information restoration via vacuum entanglement is the same as for the usual burst scenario. Thus, we conclude that vacuum entanglement does not provide a viable way of avoiding information loss for evaporating black holes.

In \cref{part}, we briefly review the notion of ``particles'' in quantum field theory and the freedom in their definition. In \cref{milne}, we consider $(1+1)$-dimensional Minkowski spacetime and show how the Minkowski vacuum can be alternatively described in terms of both Rindler particles and Milne particles. In \cref{hsusec}, we describe the key features of the moving mirror model of Hotta, Schutzhold, and Unruh \cite{hsu}. In \cref{vacent}, we give a precise description of the final ``out'' state in this model, using an unconventional notion of ``particles.'' In \cref{pecost}, we then consider the ordinary inertial particle content of the ``out'' state and show that there must be at least as many inertial particles at late times as there were Hawking particles at early times. If the mirror is brought to rest at late times, we show that the late time inertial particles must carry much more energy than the energy emitted in Hawking radiation. However, we show that if the acceleration is merely turned off at late times without changing the velocity of the mirror, then these particles will carry very little energy.
Finally, in \cref{bh} we consider an analogous vacuum entanglement of Hawking radiation particles for an evaporating black hole in $(3+1)$-dimensions. The analogous vacuum entanglement also requires the emission of at least as many inertial particles at late times as there were Hawking particles at early times, but in this case the inertial particles must be of Planck scale energy. Thus, the purification of Hawking radiation by entanglement with vacuum fluctuations in the final Minkowski region 
suffers from exactly the same problems as the usual ``final burst'' scenarios for avoiding information loss.

We restrict consideration in this paper to the theory of a free, real, massless Klein-Gordon scalar field. The scale invariance of this theory will play an essential role in our analysis.

%%%%%%%%%%%%%%%%%%%%%%%%%%%%

\section{The Definition of ``Particles'' in Quantum Field Theory}
\label{part}

Quantum field theory is---as its name suggests---the quantum theory of fields. ``Particles'' do not play any fundamental role in the formulation of quantum field theory. This fact becomes particularly evident in the study of quantum field theory in a general, nonstationary, curved spacetime, where there are many inequivalent ways to define a notion of ``particles,'' and no way appears ``preferred." Well defined physical predictions can be made by directly considering the field observables and their coupling to other systems, without introducing a notion of ``particles.''

Nevertheless, a natural and very useful notion of ``particles'' can be defined in stationary, globally hyperbolic spacetimes. In essence, for a massless Klein-Gordon field, one defines defines a ``one-particle Hilbert space'' $\mathcal H$ as the positive frequency solutions to the Klein-Gordon equation 
\ben
\nabla^a \nabla_a \phi = 0
\label{kgeq}
\een
with finite Klein-Gordon norm
\ben
||\psi||^2_{\rm KG} = 2 \, {\rm Im} \int_\Sigma \bar{\psi} n^a \nabla_a \psi \, d \Sigma
\label{kgnorm}
\een
where the bar denotes complex conjugation and the integral is taken over a Cauchy surface $\Sigma$ with unit normal $n^a$.
One can then define a Fock space ${\mathcal F}({\mathcal H})$ associated with this one-particle Hilbert space. Finally, one defines the quantum scalar field operator $\phi$ on ${\mathcal F}({\mathcal H})$ in terms of annihilation and creation operators on this Fock space. Details of this construction can be found, e.g., in section 4.3 of \cite{waldbk2}. States in ${\mathcal F}({\mathcal H})$ have a natural particle interpretation, which corresponds to the effects on a quantum mechanical system (``particle detector'') resulting from interaction with the quantum scalar field.

The requirement that the spacetime be stationary (with a globally timelike Killing field) was needed in the above construction in order to define the notion of ``positive frequency solutions'' and to ensure positivity of the Klein-Gordon norm, \cref{kgnorm}. However, the above construction will work mathematically if one can define any subspace, $\mathcal P$, of smooth, complex solutions to the Klein-Gordon equation satisfying the following properties: (i) The Klein-Gordon product \cref{kgnorm} is positive-definite on $\mathcal P$. (ii) The complex conjugate subspace $\bar{\mathcal P}$ is orthogonal to $\mathcal P$ in the Klein-Gordon product, i.e., for any $\psi_1, \psi_2 \in \mathcal P$ we have
\ben
{\rm Im}  \int_\Sigma \psi_1 n^a \nabla_a \psi_2 \, d \Sigma = 0 \, .
\een
(iii) $\mathcal P$ and $\bar{\mathcal P}$ are suitably ``complete'' in the sense that any smooth solution, $\chi$, to \cref{kgeq} with initial data of compact support can be written as $\chi = \psi_1 + \bar{\psi}_2$ with $\psi_1, \psi_2 \in \mathcal P$. Given a subspace $\mathcal P$ satisfying (i)-(iii), one can define $\mathcal H$ to be the completion of $\mathcal P$ in the Klein-Gordon norm \cref{kgnorm}. One can then define ${\mathcal F}({\mathcal H})$ and the field operator $\phi$ on ${\mathcal F}({\mathcal H})$ as in the previous paragraph. States in ${\mathcal F}({\mathcal H})$ are, of course, automatically labeled by ``particle content,'' but ``particles'' defined in this manner do not correspond to how a physical ``particle detector'' system would behave under interactions with the quantum field---except in the case of a stationary spacetime with $\mathcal P$ chosen to be the positive frequency solutions.

If one makes two different allowed choices of subspace, ${\mathcal P}_1$, ${\mathcal P}_2$, one will obtain two different constructions $({\mathcal F}({\mathcal H}_1), \phi_1)$, $({\mathcal F}({\mathcal H_2}), \phi_2)$ of Hilbert spaces and quantum field operators defined on these Hilbert spaces. One may ask whether these constructions are unitarily equivalent, i.e., whether there exists a unitary map $U : {\mathcal F}({\mathcal H}_1) \to {\mathcal F}({\mathcal H}_2)$ such that $\phi_2 = U \phi_1 U^{-1}$. If such a $U$ exists, then the constructions differ only by a labeling of states in terms of ``particles.'' If there were only finitely many linearly independent solutions of the Klein-Gordon equation (\ref{kgeq})---i.e., if $\phi$ had only a finite number of ``modes''---then unitary equivalence would always hold. However, since there are infinitely many modes of $\phi$---and thus
${\mathcal P}$ and ${\mathcal H}$ are always infinite dimensional---unitary equivalence of different constructions need not hold.

If $({\mathcal F}({\mathcal H}_1), \phi_1)$ and $({\mathcal F}({\mathcal H_2}), \phi_2)$ are unitarily equivalent, it is not difficult to explicitly solve for the unitary map $U : {\mathcal F}({\mathcal H}_1) \to {\mathcal F}({\mathcal H}_2)$ (see, e.g., section 4.4 of \cite{waldbk2}). However, even when $({\mathcal F}({\mathcal H}_1), \phi_1)$ and $({\mathcal F}({\mathcal H_2}), \phi_2)$ are not unitarily 
equivalent, one can write down a formal expression for $U \Psi_1$ for any $\Psi_1 \in {\mathcal F}({\mathcal H}_1)$. This formal expression will involve a divergent sum over infinitely many particle modes in ${\mathcal F}({\mathcal H_2})$. The divergence of this expression reflects the fact that no state in ${\mathcal F}({\mathcal H_2})$ corresponds exactly to $\Psi_1$ with respect to all field observables.

An interpretation of the formal, divergent expression for $U \Psi_1$ for the case of unitarily inequivalent constructions, $({\mathcal F}({\mathcal H}_1), \phi_1)$ and $({\mathcal F}({\mathcal H_2}), \phi_2)$, can be given as follows. Although no state in ${\mathcal F}({\mathcal H_2})$ corresponds exactly to $\Psi_1$ for all field observables, if one is interested only in a finite number of field observables, one can 
suitably {\em truncate} the divergent expression for $U \Psi_1$ to obtain a normalizable state in 
${\mathcal F}({\mathcal H_2})$ that provides an arbitrarily good approximation\footnote{For the case where the field observables are bounded operators with the structure of a $C^*$-algebra, this statement can be formulated in a precise manner via Fell's theorem \cite{fell}.} to $\Psi_1$. In the following sections, we will give formulas for a state in one representation written as a formal expression for a state in a unitarily inequivalent representation. These formulas should be interpreted in the above manner.

The fact that there are (infinitely many) inequivalent Hilbert space constructions of the theory---and, in general spacetimes, no construction appears to be ``preferred''---provides the main motivation for formulating the theory via the algebraic approach.
In the algebraic approach, one starts with an algebra of field observables and defines states as positive linear functionals on this algebra. The GNS construction then shows that every state in the algebraic sense can, if fact, be realized as a Hilbert space vector in some representation of the field algebra. Thus, the algebraic approach allows all states in all Hilbert space constructions of the theory. It enables one to formulate the theory without the awkwardness of effectively being forced to arbitrarily choose a particular Hilbert space construction from the outset. In the algebraic approach, one can see clearly that divergent expressions for states such as we will encounter below are not a difficulty of the theory but rather an artifact of trying to write a state as a vector in a Hilbert space representation to which it does not belong.

\section{Rindler and Milne Particles in $(1+1)$-Dimensional Minkowski Spacetime}
\label{milne}

Consider a massless Klein-Gordon scalar field in $(1+1)$-dimensional Minkowski spacetime, $(\mathbb R^2, \eta_{ab})$. Minkowski spacetime possesses a globally timelike translational symmetry, which can be used to define a notion of ``positive frequency solutions,'' which, in turn, can be used to define a notion of ``particles'' and a corresponding Fock space construction of the quantum field theory. In fact, however, infrared divergences arise in this construction: For test functions $f: \mathbb R^2 \to {\mathbb R}$ for which $\int d^2 x f \neq 0$, the positive frequency part of the advanced minus retarded solution with source $f$ has infinite Klein-Gordon norm, \cref{kgnorm}. We will deal with this issue, in the Fock space construction, by defining the smeared field operator, $\phi(f)$, only for test functions $f$ for which $\int d^2 x f = 0$. This restriction will be of no consequence for our considerations below.

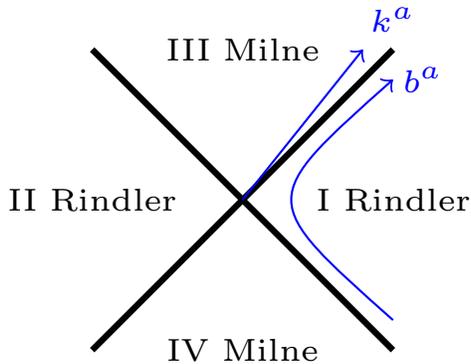
\begin{figure}[ht]
\centering
\scalebox{2}
{\begin{tikzpicture}
\draw [very thick] (-1, -1)  -- (1, 1);
\draw [very thick] (1,-1) --(-1,1);
\draw[->, blue] (1, -.8) .. controls(.1, 0) .. (1, .8);
\draw [->, blue] (0,0) -- (.8, 1);
\node at (1,0) {\tiny\textrm{I Rindler}};
\node at (-1,0) {\tiny\textrm{II Rindler}};
\node at (0,1) {\tiny\textrm{III Milne}};
\node at (0,-1) {\tiny\textrm{IV Milne}};
\node at (1.2, .8) {\tiny \color{blue} \textit{$b^a$}};
\node at (1, 1.2) {\tiny \color{blue} \textit{$k^a$}};
\end{tikzpicture}}
\caption{$(1+1)$-dimensional Minkowski spacetime, divided into the wedges I, II, III, and IV.}
\label{2dmink}
\end{figure}

Minkowski spacetime also possesses Lorentz boost symmetry. The Lorentz boost Killing field is given in terms of global inertial coordinates $(t,x)$ by
\ben
b^a = \kappa [x (\partial/\partial t)^a + t (\partial/\partial x)^a]
\een
where we have chosen to normalize $b^a$ so that $b^a b_a = -1$ on the worldline with acceleration $a = \kappa$. In terms of the null coordinates
\ben
u = t-x \,, \quad \quad \quad v = t + x
\een
we have 
\ben
b^a = \kappa [-u (\partial/\partial u)^a + v (\partial/\partial v)^a]
\een
The boost Killing field is null on the two ``Rindler horizons,'' i.e., the
two null straight lines $u=0$, $v=0$ passing through the origin. The Rindler horizons divide Minkowski spacetime into $4$ wedges, as illustrated in Fig. \ref{2dmink}. The orbits of the boost Killing field $b^a$ are future-directed timelike in ``Rindler wedge I" and are past-directed timelike in ``Rindler wedge II." We can treat each of wedge I and wedge II as a stationary, globally hyperbolic spacetime in its own right. We can thereby perform the stationary quantization construction described in the previous section in wedges I and II, using $b^a$ and $-b^a$, respectively, to define a notion of ``time translation symmetry." The notion of ``particles" thereby obtained in wedges I and II corresponds to the response of a ``particle detector'' following an orbit of $b^a$, i.e., to the particle content ``seen" by a uniformly accelerating observer.

Let ${\mathcal H}_I$ denote the one-particle Hilbert space obtained in Rindler wedge I using the boost Killing field, $b^a$, to define ``positive frequency'' there. Similarly, let ${\mathcal H}_{II}$ denote the one-particle Hilbert space obtained in Rindler wedge II using $-b^a$ (which is future directed in wedge II) to define ``positive frequency'' there. The key fact that underlies the Unruh effect is as follows: Let $f_{I\omega}$ be a wave packet in region I that is positive frequency with respect to $b^a$, with frequency peaked sharply about $\omega$. Let $f_{II\omega}$ be the wave packet in region II obtained by wedge reflection of $f_{I\omega}$ followed by complex conjugation, i.e., in wedge II (i.e., for $u > 0$ and $v < 0$) we define
\ben
f_{II \omega} (u,v) = \bar{f}_{I\omega} (-u,-v)
\een
Then $f_{II\omega}$ is positive frequency with respect to $-b^a$ in wedge II. It follows that 
\begin{equation}
F_{1\omega} = f_{I\omega} + e^{-\pi \omega/\kappa} \bar{f}_{II\omega}
\label{F1}
\end{equation}
and
\begin{equation}
F_{2\omega} = f_{II\omega} + e^{-\pi \omega/\kappa} \bar{f}_{I\omega}
\label{F2}
\end{equation}
are purely positive frequency with respect to inertial time (see, e.g., section 5.1 of \cite{waldbk2}). This implies that the Minkowski vacuum $| 0 \rangle_M$ is given in terms of Rindler particles by\footnote{The right side of \cref{mr} does not define a normalizable state in ${\mathcal F}({\mathcal H}_I) \otimes {\mathcal F}({\mathcal H}_{II})$. This formula should be interpreted in the manner described near the end of \cref{part}.} \cite{waldbk2}
\begin{equation}
| 0 \rangle_M = \Pi_i \left( \sum_n e^{-n\pi \omega_i/\kappa} |n\rangle_{iI}  |n\rangle_{iII} \right)
\label{mr}
\end{equation}
Here the product $\Pi_i$ is taken over a basis $\{f_{iI} \}$ of $\mathcal H_I$, $ |n\rangle_{iI}$ denotes the state of $n$ Rindler particles in the mode $f_{iI}$, and $|n\rangle_{iII}$ denotes the state of $n$ Rindler particles in the mode $f_{iII}$.
Thus, $| 0 \rangle_M$ consists is an entangled state of Rindler particles in wedges I and II. When restricted to a single Rindler wedge $| 0 \rangle_M$ is precisely a thermal state at temperature $T = \kappa/2 \pi$. This latter fact holds much more generally (i.e., not just for a free field), as implied by the Bisognano-Wichmann theorem \cite{bw}.

From the point of view of an inertial observer, the Rindler particles that are present in the Minkowski vacuum state $| 0 \rangle_M $ can be interpreted as a manifestation of the presence of vacuum fluctuations. The excitation of a particle detector carried by an accelerating observer would be viewed by an inertial observer as having been caused by a vacuum fluctuation. The entanglement between Rindler particles in wedges I and II seen in \cref{mr} is a manifestation of the fact that vacuum fluctuations in wedges I and II are correlated.

The above result expressing the Minkowski vacuum in terms of Rindler particles is very well known. It is much less well known---but, nevertheless, not unknown (see, e.g., section IV of \cite{hig})---that for scale invariant fields, such as the massless Klein-Gordon field, a similar quantization can be performed in wedges III and IV of \cref{2dmink}. Here we use the dilation conformal Killing field 
\ben
k^a = \kappa [t (\partial/\partial t)^a + x (\partial/\partial x)^a] = \kappa [u (\partial/\partial u)^a + v (\partial/\partial v)^a] 
\label{knorm}
\een
to define a notion of time translation symmetry in wedge III, and we use $-k^a$ in wedge IV. In this case, it is essential that the quantum field be scale invariant, since otherwise $k^a$ would not define a symmetry. We have chosen the (arbitrary) normalization of $k^a$ to correspond with our (arbitrary) normalization of $b^a$. Note that wedges III and IV are $(1+1)$-dimensional Milne universes \cite{ew}, i.e. they are FLRW models, with the orbits of $k^a$ defining the FLRW observers and with the hyperbolas orthogonal to $k^a$ defining the FLRW time slicing. The time coordinate associated with $k^a$ is the usual FLRW conformal time coordinate.

We can define the one-particle Hilbert space ${\mathcal H}_{III}$ for wedge III as the solutions in that wedge that are positive frequency with respect to $k^a$. We can similarly define ${\mathcal H}_{IV}$ for wedge IV as the solutions in that wedge that are positive frequency with respect to $-k^a$. We will refer to this quantum field construction in wedges III and IV as ``Milne quantization" and will refer to the elements of ${\mathcal H}_{III}$ and ${\mathcal H}_{IV}$ as ``Milne particles." Although ``Rindler particles'' in wedge I could be detected in a straightforward manner using a rigid particle detector carried by an accelerating observer, a particle detector of ``Milne particles'' in wedge III would need to expand with the Milne observers and be coupled to the field in a scale invariant manner.

In fact, the Milne positive frequency solutions in wedges III and IV coincide precisely with the Rindler positive frequency solutions in wedges I and II. Specifically, a left-moving Rindler particle mode in wedge I will cross the horizon at  $x=t$ with $x >0$ and become a Milne particle mode in wedge III. Similarly, a right-moving Milne particle mode in wedge IV becomes a Rindler particle mode in wedge I, a left-moving Milne particle mode in wedge IV becomes a Rindler particle mode in wedge II, and a right-moving Rindler particle mode in wedge II becomes a Milne particle mode in wedge III. These results follow immediately from the fact that, by inspection, $b^a = k^a$ when $x=t$, whereas $b^a = - k^a$ when $x= -t$. By definition, a Rindler particle mode in wedge I is a solution in wedge I that positive frequency with respect to $b^a$. A left moving Rindler particle mode in wedge I will propagate through the horizon $x=t$, $x >0$ that forms the boundary between wedges I and III. On the horizon, the mode will be positive frequency with respect to $b^a$. But since $k^a = b^a$ on the horizon, it is also positive frequency with respect to $k^a$ on the horizon. It then follows immediately that the solution will be positive frequency with respect to $k^a$ throughout wedge III, as claimed. The other claims follow similarly. Thus, if we decompose the one-particle Hilbert spaces ${\mathcal H}_{I}$, ${\mathcal H}_{II}$, ${\mathcal H}_{III}$, and ${\mathcal H}_{IV}$ into their left and right moving parts, we have
\ben
{\mathcal H}_{IL} \equiv {\mathcal H}_{IIIL} \, , \quad {\mathcal H}_{IR} \equiv {\mathcal H}_{IVR} \, , \quad {\mathcal H}_{IIL} \equiv {\mathcal H}_{IVL} \, , \quad {\mathcal H}_{IIR} \equiv {\mathcal H}_{IIIR} \, .
\label{hsr}
\een

It follows immediately from \cref{hsr} that the relationship between inertial positive frequency solutions and positive frequency solutions with respect to ``Milne time'' in wedges III and IV is exactly the same as the relationship between inertial positive frequency solutions and positive frequency solutions with respect to Rindler time in wedges I and II. Consequently, using \cref{mr}, we can immediately write down the following expression for the Milne particle content of the Minkowski vacuum 
\begin{equation}
| 0 \rangle_M = \Pi_i \left( \sum_n e^{-n\pi \omega_i/\kappa} |n\rangle_{iIII}  |n\rangle_{iIV} \right)
\label{mm}
\end{equation}
Here $| n \rangle_{iIII}$ is a state of $n$ Milne particles in wedge III in mode $f_{iIII}$ that is sharply peaked near frequency $\omega_i$, whereas $| n \rangle_{iIV}$ is the corresponding state of $n$ Milne particles in wedge IV in the mode, $f_{iIV}$, obtained by reflection of $f_{iIII}$ through the origin followed by complex conjugation. Again, the product over a basis of such modes is taken.
Thus, $| 0 \rangle_M$ is an entangled state of Milne particles in wedges III and IV. When restricted to a single Milne wedge, $| 0 \rangle_M$ is precisely a thermal state of Milne particles at temperature $T = \kappa/2 \pi$. This result also is a consequence of the Hislop-Longo theorem \cite{hl}. As in the Rindler case, from the point of view of an inertial observer, the Milne particles that are present in the Minkowski vacuum state $| 0 \rangle_M $ would be interpreted as a manifestation of the presence of vacuum fluctuations. The entanglement of the Milne particles in wedges III and IV seen in \cref{mm} would be interpreted as a manifestation of the correlations in vacuum fluctuations in wedges III and IV.

Using \cref{hsr}, we may also write \cref{mr} in terms of right and left moving particles as
\begin{equation}
| 0 \rangle_M = \Pi_{i,j} \left( \sum_n e^{-n\pi \omega_i/\kappa} |n\rangle_{iIIIR}  |n\rangle_{iIR} \right) \left( \sum_m e^{-m\pi \omega_j/\kappa} |m\rangle_{jIIIL}  |m\rangle_{jIIL} \right)
\label{mmr}
\end{equation}
This shows that in the Minkowski vacuum, the right moving Milne particles in region III are entangled with corresponding right moving Rindler particles in region I (across the horizon $x=t$), whereas left moving Milne particles in region III are entangled with left moving Rindler particles in region II (across the horizon $x=-t$). 

Finally, we comment that for Minkowski spacetime of dimension $d > 2$, the boost Killing field $b^a$ is timelike in the wedges I and II defined by the null {\em planes} $t = \pm x$. Rindler quantization can again be carried out in these wedges, and the Minkowski vacuum again can be expressed in terms of Rindler particles by \cref{mr}. On the other hand, the dilation conformal Killing field $k^a$ is timelike in the future and past null {\em cones} $t = \pm r$. For a massless scalar field, Milne quantization can again be carried out in these future and past light cones, and the Minkowski vacuum can again be expressed in terms of Milne particles by \cref{mm}. However, it is only in $1+1$ dimensions that the null planes and null cones coincide, so that the Rindler and Milne particles can be identified. Thus, although eqs.~(\ref{mr}) and (\ref{mm}) hold in Minkowski spacetime of any dimension, eqs.~(\ref{hsr}) and (\ref{mmr}) hold only in $(1+1)$-dimensions.

\section{The Moving Mirror Spacetime of Hotta, Schutzhold, and Unruh}
\label{hsusec}

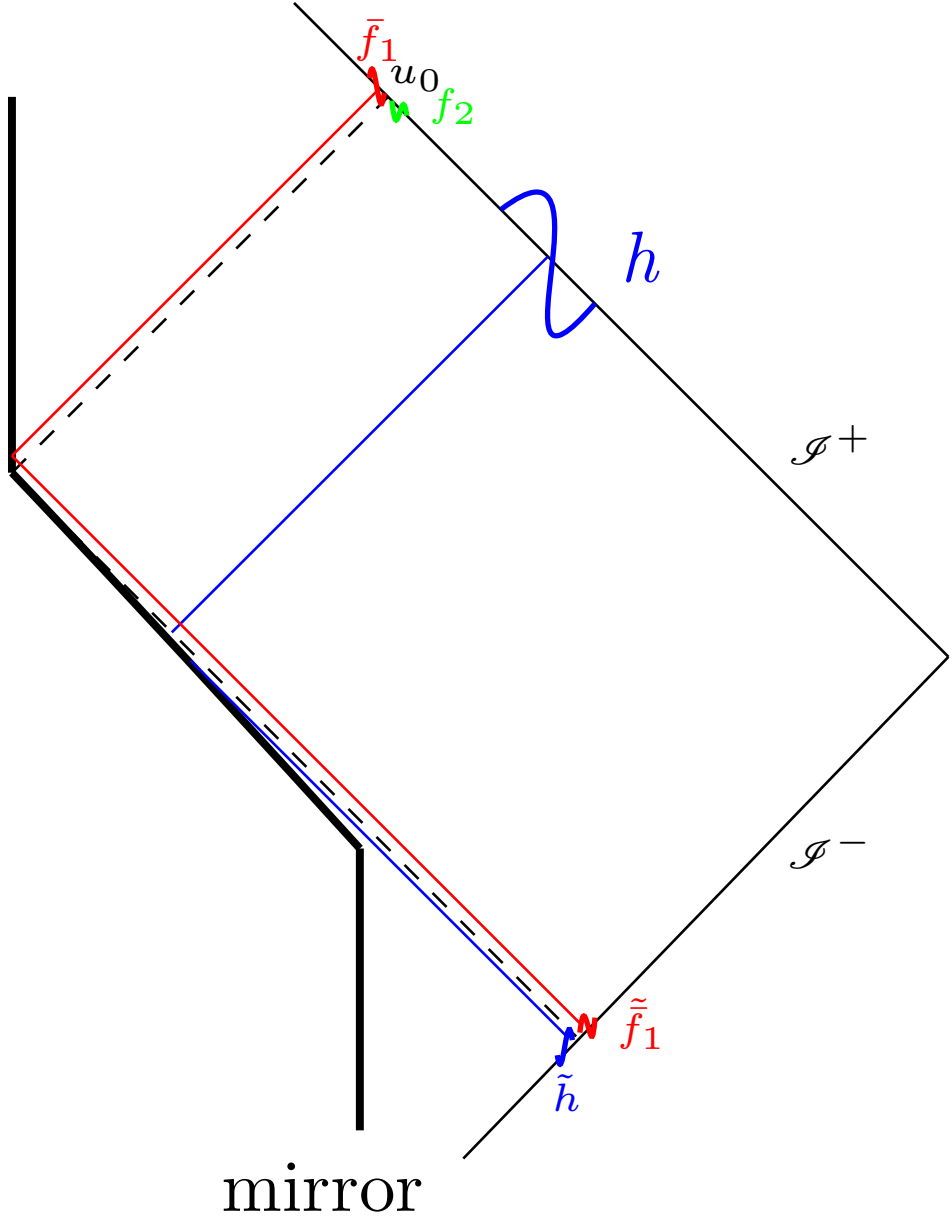
\begin{figure}[ht]
\centering
\scalebox{2.5}{\begin{tikzpicture}
\draw [very thick] (1,-2.5)  -- (1, -1);
\draw [very thick] (1,-1) .. controls (0,.1) and (-.47,.6) .. (-.85,1);
\draw [very thick] (-.85,1) -- (-.85, 3);
\draw [dashed] (-.85, 1) -- (2.15, -2);
\draw [dashed] (-.85, 1) -- (1.15, 3);
%\draw (1.55, -2.70) -- (4.15, 0);
\draw (1.55, -2.65) -- (4.13, .02);
%\draw (.65, 3.5) -- (4.15, 0);
\draw (.65, 3.5) -- (4.13, .02);
\draw [blue, thick] (1.75, 2.4) .. controls (2.4, 2.9) and (1.65, 1.20) .. (2.25, 1.9);
\draw [blue, thick] (2.05, -2.1) .. controls (2.08, -2.3) and (2.10, -1.8) .. (2.13, -2.02);
\draw [red, thick] (2.17, -1.98) .. controls (2.2, -1.7) and (2.23, -2.2) .. (2.25, -1.9);
\draw [red, thick] (1.13, 3.02) .. controls (1.1, 2.8) and (1.08, 3.3) .. (1.05, 3.1);
\draw [green, thick] (1.17, 2.98) .. controls (1.2, 2.7) and (1.22, 3.1) .. (1.25, 2.9);
\draw [blue] (2, 2.15) -- (0, .15);
\draw [blue] (.1, 0) -- (2.1, -2) ;
\draw [red] (2.17, -1.93) -- (-.85, 1.09);
\draw [red] (-.85, 1.09) -- (1.1, 3.04);
\node at (.8,-2.8) {\textrm{mirror}};
\node at (2.5, 2.15) {\color{blue}\textit{h}};
\node at (1.1, 3.3) {\tiny\color{red}\textit{$\bar{f}_1$}};
\node at (1.5, 2.95) {\tiny\color{green}\textit{$f_2$}};
\node at (2.5, -1.93) {\tiny\color{red}\textit{$\tilde{\bar{f}}_1$}};
\node at (2.1, -2.3) {\tiny \color{blue}\textit{$\tilde{h}$}};
\node at (1.3, 3.10) {\tiny\textit{$u_0$}};
\node at (3.5, 1.15) {\tiny\textit{$\mathscr I^+$}};
\node at (3.5, -1.0) {\tiny\textit{$\mathscr I^-$}};
%\node at (3.3, 1.3) {\tiny\textit{$u = 0$}};
%\draw (3, 1) -- (3.13, 1.13);
\end{tikzpicture}}
\caption{The moving mirror spacetime of Hotta, Schutzhold, and Unruh (see text).}
\label{mmspacetime}
\end{figure}

We turn now to consideration of the moving mirror spacetime considered by Hotta, Schutzhold, and Unruh \cite{hsu}. The set-up is illustrated in \cref{mmspacetime}. The thick black line represents the trajectory of a mirror in $(1+1)$-dimensional Minkowski spacetime. The motion of the mirror can be divided into the following 3 eras: 

\begin{enumerate}
\item At early times, $t < - C$ (with $C> \kappa $, with $\kappa$ being the constant appearing in \cref{acc} below), the mirror is at rest at $x=0$.
\item At $u= -C$, the mirror begins to accelerate to the left. After a transition period (shown as a sharp corner in the figure, but the actual transition is assumed to be smooth), the mirror follows a trajectory given in terms of the null coordinates $v = t + x$ and $u = t - x$ by
\begin{equation}
 v = - \frac{1}{\kappa} e^{-\kappa u} \, .
 \label{acc}
\end{equation}
It follows this trajectory starting at retarded time $u= 0$ and continuing until $u=u_1$, where $\kappa u_1 \gg 1$. Note that this trajectory asymptotes to the null line $v=0$.
\item Beginning at $u=u_1$, the mirror begins transitioning back to inertial motion, and it is assumed to have become inertial by the retarded time $u=u_0$ at which the mirror trajectory intersects the null line $v=0$. (The line $v=0$ is shown in \cref{mmspacetime} as the black dashed line with slope $-1$; the line $u=u_0$ is shown as the black dashed line with slope $+1$.) In \cref{mmspacetime} the mirror is shown as returning to rest at late times, but all that is necessary is that the mirror be inertial for $v > 0$. Indeed, a case of interest in our analysis will be one where the velocity of the mirror remains very large at late times. The return to inertial motion is assumed to be smooth, even though it is depicted as a sharp transition in \cref{mmspacetime}.
\end{enumerate}

A massless quantum scalar field is assumed to be present in the spacetime to the right of the mirror and to satisfy Dirichlet boundary conditions at the mirror. In the initial period, where the mirror is at rest, the scalar field is assumed to be in its ground/vacuum state for the Minkowski half-space with a static mirror. Our analysis of the behavior of the scalar field in this spacetime will be based upon the following two key observations:

\begin{itemize}
\item By assumption, the mirror is in inertial motion for $u \geq u_0$. Consider the spacetime region $u>u_0$ , i.e., the region  lying above the black dashed line with slope $+1$ shown in \cref{mmspacetime}. We claim that the state of the quantum field in this region is identical to that of the ground/vacuum state for the given inertial motion of the mirror in this region. To see this, we note that the backwards in time propagation of the field starting from any event in the region $u > u_0$ is determined by the past-directed null geodesics starting in this region. The null geodesics that move towards the right (going backward in time) go directly to past null infinity. The null geodesics that move towards the left (going backwards in time) bounce off the mirror when it is in its final inertial state and then go to past null infinity. Thus, none of the null geodesics starting in the region $u > u_0$ ``see'' the non-inertial motion of the mirror. If follows that the quantum field $\phi$ also does not ``see'' the non-inertial motion, i.e., all of the correlation functions  $\langle \phi(x_1) \dots \phi(x_n) \rangle$ of the quantum field in the region $u > u_0$ are identical to the correlation functions that would occur if the mirror was in its final inertial motion at all times.

\item Consider a wavepacket $h(u)$ at future null infinity of (inertial) positive frequency peaked near $\omega$ that emerges at a retarded time corresponding to the era when the mirror undergoes the motion \cref{acc}, as illustrated in \cref{mmspacetime}.
Propagate this wavepacket backwards in time. This wavepacket will ``bounce off'' of the mirror and become a wavepacket with time dependence at past null infinity that goes as 
\ben
\tilde{h}(v) \sim \exp(-i \omega u(v)) \sim \exp[i \frac{\omega}{\kappa} \ln(-\kappa v)] \, .
\label{th}
\een
Thus, the blueshift of the wavepacket resulting from its reflection by the mirror is the same as would occur if the wavepacket propagated near the horizon of a black hole of surface gravity $\kappa$ in the Hawking effect---which is why the mirror motion \cref{acc} was chosen. Thus, by the same analysis as originally given for the Hawking effect \cite{hawk}, \cite{wald1}, the mirror will emit Hawking radiation to $\mathscr I^+$ during the era $0 < u < u_1$ where its motion is given by \cref{acc}. Note that if we identify the region $v < 0$ of $\mathscr I^-$ in \cref{mmspacetime} with the portion $v<0$ of $\mathscr I^-$ of  \cref{2dmink} corresponding to the Milne wedge IV, then for the normalization of $k^a$ given by \cref{knorm}, $\tilde{h}(v)$ has frequency $\omega$ with respect to Milne time $-\kappa^{-1} \ln(-v)$.
\end{itemize}

\section{Vacuum Entanglement of Hawking Radiation in the Moving Mirror Spacetime}
\label{vacent}

We now consider particle creation in the spacetime of \cref{mmspacetime}. To do so, we need definitions of ``ingoing particles'' at past null infinity, $\mathscr I^-$, and ``outgoing particles'' at future null infinity, $\mathscr I^+$. We have a natural notion of (inertial) time translation symmetry at both $\mathscr I^-$ and $\mathscr I^+$ and, of course, it would be completely standard to use the positive frequency solutions with respect to inertial time at $\mathscr I^-$ and $\mathscr I^+$ to define ingoing and outgoing particle states. We will indeed use the standard definition of ingoing particles at $\mathscr I^-$, i.e., we define $\mathcal P_{\rm in}$ (see \cref{part}) to be the subspace of solutions that are positive frequency with respect to $v$ at $\mathscr I^-$, and we define $\mathcal H_{\rm in}$ to be the corresponding one-particle Hilbert space. In the next section, we will be concerned with the inertial particle content of the outgoing state, and in that section we will define $\mathcal P_{\rm out}$ and $\mathcal H_{\rm out}$ in the usual manner in terms of the solutions that are positive frequency with respect to $u$ at $\mathscr I^+$. However, in this section, it will be very useful for analyzing the entanglement of the Hawking radiation emitted by the mirror to use a non-standard definition, $\mathcal H'_{\rm out}$, of outgoing particles at $\mathscr I^+$. Basically, we wish to define $\mathcal H'_{\rm out}$ by (i) using the usual inertial notion of particles---i.e., positive frequency solutions with respect to $u$---at early and intermediate times with $u \ll u_0$; (ii) using the Rindler notion of particles---i.e., positive frequency with respect to $-\kappa^{-1}\ln|u-u_0|$---for $u < u_0$ with $u \sim u_0$; and (iii) using the Milne notion of particles---i.e., positive frequency with respect to $\kappa^{-1}\ln(u-u_0)$---at late times, $u > u_0$. We cannot do this literally/exactly because any solution that is positive frequency with respect to $u$ cannot vanish in any open region in $u$ and thus cannot vanish identically in the region where we wish to use the Rindler and Milne notion of particles. Nevertheless, we can define $\mathcal H'_{\rm out}$ to be given by the subspace, $\mathcal P'_{\rm out}$, of solutions (see \cref{part}) spanned by: (a) Wavepackets that are positive frequency with respect to inertial time, $u$, that reach $\mathscr I^+$ at $u \ll u_0$ and have negligible ``tails'' extending to $u \gtrsim u_0$. (We may then slightly modify these wavepackets to make them strictly vanish for $u \gtrsim u_0$.) (b) Wavepackets that reach $\mathscr I^+$ at times near $u_0$ with $u < u_0$ and are purely positive frequency with respect to Rindler time, $-\kappa^{-1}\ln|u-u_0|$. (c) Wavepackets that reach $\mathscr I^+$ at times $u >u_0$ and are purely positive frequency with respect to Milne time, $\kappa^{-1}\ln(u-u_0)$. (d) An arbitrary choice of additional solutions needed to ``complete'' $\mathcal P'$ so that it satisfies conditions (i)-(iii) of \cref{part}. Thus, as desired, elements of $\mathcal H'_{\rm out}$ correspond to ordinary inertial particles for $u \ll u_0$, correspond to Rindler particles for $u \sim u_0$ with $u < u_0$, and correspond to Milne particles for $u > u_0$. Note that since we may take $u_0$ to be arbitrarily large, the regime $u \ll u_0$ may be assumed to include an arbitrarily long era where the mirror undergoes acceleration given by \cref{acc}.

We now calculate the particle creation occurring in the spacetime of \cref{mmspacetime}. Nontrivial particle creation with respect to our notion of ``in'' and ``out'' particles starting from $|0 \rangle_{\rm in}$ occurs if a solution in $\mathcal P_{\rm in}$ does not lie entirely in $\mathcal P'_{\rm out}$, i.e., if, after evolution to $\mathscr I^+$, it has a part lying in $\bar{\mathcal P}'_{\rm out}$. We are primarily concerned with the properties of the Hawking radiation emitted at $0 < u \ll u_0$ while the mirror undergoes the motion \cref{acc}. To analyze this, we proceed in the same manner as used to analyze particle creation by black holes \cite{hawk}, \cite{wald1}: We start with a normalized (i.e., unit Klein-Gordon norm) positive frequency wavepacket, $h(u)$, at $\mathscr I^+$ that is composed of frequencies near $\omega$ and that is localized in time during the ``intermediate era'' where \cref{acc} holds, but with $u \ll u_0$. We propagate this wavepacket backwards into the past, where it reflects off of the mirror and evolves to the wavepacket $\tilde{h} (v)$ at $\mathscr I^-$, as illustrated in \cref{mmspacetime}. As already noted above, the wavepacket $\tilde{h}(v)$ will have time dependence given by \cref{th}, and thus will be purely positive frequency (with frequencies peaked near $\omega$) with respect to Milne time in the region $v < 0$. The wavepacket $\tilde{h}(v)$ will not be positive frequency with respect to $v$. However, define the wavepacket $\tilde{f}_1(v)$ on $\mathscr I^-$ by reflecting $\tilde{h}(v)$ about $v=0$ and then taking its complex conjugate
\ben
\tilde{f}_1(v) = \bar{\tilde{h}}(-v)
\label{tf1}
\een
Then $\tilde{f}_1$ is purely positive frequency with respect to Rindler time for $v > 0$, and will have unit Klein-Gordon norm. By exactly the same calculation as led to eqs~(\ref{F1}) and (\ref{F2}) above, we find that the quantities
\ben
\tilde{F}_1(v) = \tilde{f}_1(v) + e^{-\pi \omega/\kappa} \bar{\tilde{h}}(v)
\label{tF1}
\een
and 
\ben
\tilde{F}_2(v) = \tilde{h}(v) + e^{-\pi \omega/\kappa} \bar{\tilde{f_1}}(v)
\label{tF2}
\een
are purely positive frequency with respect to inertial time $v$. 

If we propagate the solution given by the data $\tilde{h}(v)$ at $\mathscr I^-$ forward in time, we will, of course, get back the original wavepacket $h(u)$ at $\mathscr I^+$. This outgoing solution lies in $\mathcal P'_{\rm out}$. If we propagate the solution given by $\tilde{f}_1(v)$ at $\mathscr I^-$ forward in time, it will reflect off the mirror while the mirror is in inertial motion, given by
\ben
v = \alpha(u-u_0)
\een
where $\alpha = (1+V)/(1-V)$, where $V$ is the final velocity of the mirror.
The resulting wavepacket $f_1(u)$, at $\mathscr I^+$ will therefore have time dependence of the form
\ben
f_1(u) = -\bar{\tilde{h}}(-v(u)) \sim \exp[-i \omega/\kappa \ln(\kappa \alpha (u-u_0))] \, .
\een
Thus, $f_1$ is purely positive frequency---with frequencies peaked at $\omega$---with respect to Milne time $\kappa^{-1} \ln (u-u_0)$ in the region $u > u_0$. Thus, $f_1$ also lies in $\mathcal P'_{\rm out}$. Note that by conservation of the Klein-Gordon inner product, the wavepacket $f_1$ will have unit Klein-Gordon norm.

Let us now evolve the initial positive frequency solutions $\tilde{F}_1$ and $\tilde{F}_2$ from $\mathscr I^-$ to $\mathscr I^+$. The solution $\tilde{F}_1$ evolves to $f_1(v) + e^{-\pi \omega/\kappa} \bar{h}(v)$. These two terms give, respectively, its decomposition into $\mathcal P'_{\rm out}$ and $\bar{\mathcal P}'_{\rm out}$. Similarly, $\tilde{F}_2$, evolves to $h(v) + e^{-\pi \omega/\kappa} \bar{f}_1(v)$, which also provides its decomposition into $\mathcal P'_{\rm out}$ and $\bar{\mathcal P}'_{\rm out}$. It follows immediately by the same type of analysis as used to calculate particle creation by black holes \cite{wald1}
that the ``out'' state $\Psi \in {\mathcal F} ({\mathcal H}_{\rm out})$ takes the form
\begin{equation}
\Psi =  \left(\sum_n e^{-n\pi \omega/\kappa} |n\rangle_{h}  |n\rangle_{f_1} \right) \otimes \Psi'
\label{hm}
\end{equation}
Here $ |n\rangle_{h}$ is a outgoing state of $n$ inertial particles in the Hawking mode $h(u)$, $|n\rangle_{f_1}$ is an outgoing state of $n$ Milne particles in the mode $f_1(u)$, and $\Psi'$ describes the state of the system with respect to the modes that are orthogonal to 
both $h$ and $f_1$ in our one-particle ``out'' Hilbert space ${\mathcal H}_{\rm out}$. \Cref{hm} shows that the Hawking particles that are emitted at $u \ll u_0$ are entangled with outgoing Milne particles in the region $u > u_0$. These Milne particles are the ``partner particles'' of the Hawking radiation particles in the sense of \cite{hsu} (see also \cite{tyh}). Tracing \cref{hm} over the Milne particle degrees of freedom, we see that the Hawking particles are described by a precisely thermal density matrix at $T = \kappa/2 \pi$. Tracing \cref{hm} over the Hawking particle degrees of freedom, we see that the Milne particles also are in precisely a thermal state at $T= \kappa/2\pi$. As we saw in \cref{milne}, this fact is compatible with the fact that the state of the scalar field for $u > u_0$ is the ground/vacuum state for the final inertial motion of the mirror.

In the moving mirror spacetime of \cref{mmspacetime}, there is no black hole or singularity of any kind present, so there can be no ``information loss.''  As in the black hole case, the Hawking radiation is emitted in a mixed thermal state. As seen from \cref{hm}, the ``information'' needed to purify the Hawking radiation emerges to infinity in the form of Milne particles at late times ($u > u_0$), i.e., after the mirror has returned to inertial motion. 
As discussed in \cref{milne}, from the inertial point of view, a thermal distribution of Milne particles corresponds to vacuum fluctuations. Thus, \cref{hm} can be interpreted as saying in a precise sense that the Hawking radiation is entangled with vacuum fluctuations, i.e. {\em in the moving mirror spacetime of \cref{mmspacetime}, the information needed to purify the Hawking radiation is contained in the vacuum fluctuations of the quantum field that are present in the late time region $u > u_0$.}

\section{The Inertial Particle and Energy Cost of Vacuum Entanglement}
\label{pecost}

The discussion at the end of the previous section may suggest that the purification of the Hawking radiation in the moving mirror spacetime \cref{mmspacetime} is ``cost-free'' with regard to (inertial) particles and energy. The vacuum fluctuations in the region $u > u_0$ provide an infinite reservoir of ``information'' to entangle with the Hawking radiation. Yet, this vacuum region contains no energy and no apparent inertial particle content! Thus, it might appear that the Hawking radiation can be purified without requiring a ``final burst'' of inertial particles and without requiring the emission of energy.

However, we will now show that, in fact, there is a large cost in inertial particle emission. Specifically, the number inertial particles that must be emitted near $u \sim u_0$ is at least as great as the total number of Hawking particles that were emitted during the phase where the mirror motion was given by \cref{acc}. The basic reason why this is so can be understood as follows: In the outgoing state $|\Psi \rangle$ of \cref{hm}, the Milne particles in mode $f_1$ are entangled with the Hawking particles in mode $h$. However, according to \cref{mmr}, in the vacuum state, the Milne particles in mode $f_1$  would be entangled with corresponding {\em Rindler} particles in mode $f_2$---where $f_2$ is obtained by reflecting $f_1$ on $\mathscr I^+$ about $u = u_0$ and then taking its complex conjugate, i.e.,
\ben
f_2(u-u_0) = \bar{f}_1 (u_0 - u) \, .
\een
But the Rindler particles in mode $f_2$ cannot be entangled with the Milne $f_1$-particles since the $f_1$-particles are already entangled with Hawking particles. Thus, the Rindler particles in mode $f_2$ cannot be in a state where they can be interpreted as vacuum fluctuations. Inertial particles must be present!

The considerations of the previous paragraph can be made precise as follows. We are interested in the inertial particle content of the outgoing state $|\Psi \rangle$ of \cref{hm}. Thus, we wish to express $|\Psi \rangle$ as a state in ${\mathcal F}(\mathcal H_{\rm out})$, where $\mathcal H_{\rm out}$ is the usual one-particle Hilbert space of solutions that are positive frequency with respect to inertial time $u$ at $\mathscr I^+$. The relationship between ${\mathcal F}(\mathcal H_{\rm out})$ and ${\mathcal F}(\mathcal H'_{\rm out})$ is given by a Bogoliubov transformation. A key observation concerning this Bogoliubov transformation is that---by the same calculation as in eqs.~(\ref{F1})-(\ref{F2}) and eqs.~(\ref{tF1})-(\ref{tF2})---the ``late time'' outgoing solution\footnote{We have inserted the factor of $[1 - \exp(-2 \pi \omega/\kappa)]^{-1/2}$ in \cref{Fout} so that $F_1$ will have unit Klein-Gordon norm.}
\ben
F_1(u) = \frac{f_1(u) + e^{-\pi \omega/\kappa} \bar{f}_2 (u)}{\sqrt{1 - e^{-2 \pi \omega/\kappa}}}
\label{Fout}
\een
is purely positive frequency with respect to inertial time. Let $a, a^\dagger$ denote the annihilation and creation operators on ${\mathcal F}(\mathcal H_{\rm out})$ and let $b, b^\dagger$ denote the annihilation and creation operators on ${\mathcal F}(\mathcal H'_{\rm out})$. \Cref{Fout} implies that
\begin{equation}
a(F_1) = \frac{b(f_1) - e^{-\pi \omega/\kappa} b^\dagger ({f}_2)}{\sqrt{1-e^{-2\pi \omega/\kappa}}}
\end{equation}
Consequently, the expected number of outgoing {\em inertial} particles in mode $F_1$ is given by
\begin{eqnarray}
\langle N (F_1) \rangle = \langle \Psi| a^\dagger (F_1) a(F_1)| \Psi \rangle &=& \frac{1}{1-e^{-2\pi \omega/\kappa}}\big[\langle \Psi| b^\dagger (f_1) b(f_1) - e^{-\pi \omega/\kappa} b^\dagger (f_1) b^\dagger ({f}_2) \nonumber \\
&-& e^{-\pi \omega/\kappa} b(f_1)b({f}_2) + e^{-2\pi \omega/\kappa} b(f_2)b^\dagger({f}_2)| \Psi \rangle \big]
\label{N1}
\end{eqnarray}
From the form of $\Psi$, \cref{hm}, it follows immediately that $\langle \Psi | b^\dagger (f_1) b^\dagger ({f}_2) | \Psi \rangle = \langle \Psi | b(f_1) b ({f}_2) | \Psi \rangle = 0$, i.e., the middle two terms on the right side of eq.~(\ref{N1}) vanish. The last term on the right side of eq.~(\ref{N1}) is manifestly positive. Thus, we obtain
\begin{equation}
\langle N(F_1) \rangle > \frac{1}{1-e^{-2\pi \omega/\kappa}} \langle \Psi| b^\dagger (f_1) b(f_1)| \Psi \rangle > \langle N(f_1) \rangle= \langle N(h)\rangle
\end{equation}
Here $\langle N(f_1) \rangle$ is the expected number of Milne particles in mode $f_1$ in the outgoing state $\Psi$ of \cref{hm}, and the last equality reflects the fact that this is equal to the expected number of Hawking particles in mode $h$. Thus, the expected number of {\em inertial} particles emitted in mode $F_1$ always is larger than the expected number of Hawking particles emitted in mode $h$. Since this is true for all Hawking particles---or, at least, those emitted at $u \ll u_0$---{\em the total number of non-Hawking inertial particles emitted at late times must be greater than the total number of Hawking particles emitted.}

We now consider the energy cost of the emission of these late time inertial particles. Since the mode $F_{i1}$ associated with the Hawking mode $h_i$ is not an eigenstate of inertial energy and since the different $F_{i1}$ modes may overlap at $\mathscr I^+$, we do not know of a simple way to obtain a rigorous lower bound on the total energy $E_B$ associated with late time emission. Nevertheless, if we let $e(F_{i1})$ denote the classical energy of the mode $F_{i1}$ and we let $N(F_{i1})$ denote the expected number of particles emitted in mode $F_{i1}$, then the formula
\begin{equation}
E_B \sim \sum_i N(F_{i1}) e(F_{i1})
\label{peest}
\end{equation}
should provide a reasonable estimate of the non-Hawking emitted energy. Since we already know that there are at least as many $F_{i1}$-particles as Hawking particles in mode $h_i$, the key issue is how large $e(F_{i1})$ is. This can be estimated as follows:

Consider a Hawking mode $h_i(u)$ that is localized near retarded time $u_i$ at $\mathscr I^+$, where $0 < u_i < u_1$, so the mirror motion is given by \cref{acc} at this retarded time. When propagated backwards into the past, this Hawking mode will bounce off the mirror near advanced time $v_i = - 1/\kappa \exp(-\kappa u_i)$. The (forward in time) velocity of the mirror at this time is
\ben
V_i = -\frac{1-e^{-\kappa u_i}}{1+e^{-\kappa u_i}}
\een
After it bounces off the mirror in its backward in time evolution, the frequency of the wavepacket with respect to inertial time will be blueshifted by the factor 
\ben
\frac{1 - V_i}{1 + V_i} = \exp(\kappa u_i)
\een
Thus, the wavepacket $\tilde{h}_i(v)$ at $\mathscr I^-$ will be composed of inertial frequencies peaked around 
\ben
\tilde{\omega}_i = \omega_i \exp(\kappa u_i)
\label{tom}
\een
where $\omega_i$ is the peak frequency of the original Hawking wavepacket $h_i(u)$ at $\mathscr I^+$. The wavepacket $\tilde{f}_{i1}(v)$ at $\mathscr I^-$ will therefore also have inertial frequencies peaked about $\tilde{\omega}_i$. 

Now, suppose that the mirror returns to rest at $u = u_0$. Then when the wavepacket $\tilde{f}_1(v)$ is propagated forward and bounces off the mirror at $u > u_0$, there will be no change in the inertial frequencies of this wavepacket. Consequently, the Milne particle wavepacket $f_{i1}(u)$ at $\mathscr I^+$ will also have inertial frequencies peaked about $\tilde{\omega}_i$, \cref{tom}. The Rindler particle wavepacket $f_{i2}(u)$ will also have inertial frequencies peaked about $\tilde{\omega}_i$. Thus, the inertial mode $F_{i1}$ will have energy
\ben
e(F_{i1}) \sim \tilde{\omega}_i = \omega_i \exp(\kappa u_i)
\label{paren}
\een
For $\kappa u_i \gg 1$, this is enormously greater than the energy, $\omega_i$, of the original Hawking mode $h_i(u)$. Thus, if the mirror returns to rest at $u=u_0$, there will be an enormous burst of energy emitted at $u$ near $u_0$ associated with the purification of the Hawking radiation. This burst of energy is much {\em greater} than the total energy emitted in Hawking radiation during the entire era where the mirror undergoes the motion \cref{acc}.

However, suppose that instead we simply turn off the acceleration at some time $u_1$ with $u_1 < u_0$ (but with $\kappa u_1 \gg 1$ as we have been assuming), e.g., suppose that for $0 < u < u_1$ the mirror motion is given by \cref{acc}, but for $u > u_1$ the mirror motion is given by
\ben
v = - \frac{1}{\kappa} e^{-\kappa u_1} + e^{-\kappa u_1}(u-u_1)
\label{mna}
\een
Then the final velocity, $V_f$ of the mirror will be
\ben
V_f = -\frac{1-e^{-\kappa u_1}}{1+e^{-\kappa u_1}}
\een
Consequently, when the wavepacket $\tilde{f}_{i1}(v)$ is propagated forward in time and bounces off the mirror at $u > u_0$, its inertial frequencies will be redshifted by the factor $\exp(-\kappa u_1)$. Thus, the Milne particle wavepacket $f_{i1}(u)$ at $\mathscr I^+$ will have inertial frequencies peaked about  $\omega_i \exp[-\kappa (u_1-u_i)]$. In this case, the inertial particle mode $F_{i1}$ will have energy
\ben
e(F_{i1}) \sim \omega_i \exp[-\kappa (u_1- u_i)]
\een
Consequently, the energy required to purify the Hawking radiation in this case is much {\em less} than total energy emitted in the Hawking radiation, i.e., it should be possible to purify the Hawking radiation at a negligible energy cost.

It should be noted that the above redshifts and blueshifts also have a corresponding effect on the time spread of the wavepackets. If the Hawking wavepacket $h_i(u)$ has time spread $\Delta u_i \gtrsim 1/\omega_i$ about $u_i$ at $\mathscr I^+$, then the corresponding wavepacket $\tilde{h}_i$ at $\mathscr I^-$ will have spread of order $\Delta u_i \exp(-\kappa u_i)$ about $v_i = -\kappa^{-1} \exp(-\kappa u_i)$. In the case where the mirror returns to rest for $u > u_0$, it follows that the mode $F_{i1}$ at $\mathscr I^+$ will be peaked near $u-u_0 = \pm \kappa^{-1} \exp(-\kappa u_i)$ with spread of order $\Delta u_i \exp(-\kappa u_i)$. Thus, assuming that $\kappa u_i \gg 1$, the purification of the Hawking radiation occurs in a ``final burst'' that is highly localized near $u=u_0$. However, in the case where the mirror follows the trajectory \cref{mna} for $u > u_0$, the mode $F_{i1}$ at $\mathscr I^+$ will have a time spread of order $\Delta u_i \exp[\kappa (u_1- u_i)]$, so the purification of the Hawking radiation will occur over a very long period of time. It also should be noted that in this case the modes $F_{i1}(u)$ will be sufficiently spread out in time that they will significantly overlap with each other\footnote{Indeed, they may even overlap with the early time Hawking emission, resulting in possible inconsistencies in our definition of $\mathcal H'_{\rm out}$}, so interference effects may be important and the estimate \cref{peest} may not be reliable.

The above conclusions on the energy cost of purifying the Hawking radiation can be confirmed from the formula for energy emission associated with moving mirror motion. For a mirror moving on the general trajectory 
\ben
v = p(u)
\een
the energy flux to $\mathscr I^+$ is given by \cite{df}
\ben
T_{uu} = \frac{1}{16 \pi} \left[ \left( \frac{p''}{p'} \right)^2 - \frac{2}{3} \frac{p'''}{p'} \right]
\label{mse}
\een
where the primes denote derivatives with respect to $u$.
During the era where the mirror trajectory is given by \cref{acc}, the energy flux is given by 
\ben
T_{uu} = \frac{1}{48 \pi} \kappa^2
\een
which corresponds to a thermal energy flux of Hawking radiation at temperature $T = \kappa/2 \pi$. Since the mirror follows the trajectory \cref{acc} for $0 < u < u_1$, the total energy radiated in Hawking radiation is
\ben
E_H = \frac{1}{48 \pi} \kappa^2 u_1
\label{thf}
\een

We can return the mirror to rest at $u=u_0$, $v=0$ by having it follow the trajectory\footnote{We chose the form of $p(u)$ in \cref{mtr} because extremization of \cref{mse} suggests that the motion that minimizes the total energy emission should be such that $p'(u)$ depends exponentially on $u$. We believe that \cref{ebrest} below should provide a good estimate of the minimum energy emission needed to return the mirror to rest at $v=0$, but we have not attempted to prove this.}
\ben
p(u) = -\left( \frac{1}{\kappa} + \frac{1}{\alpha} \right) e^{-\kappa u_1} + e^{-\kappa u_1} \frac{1}{\alpha}e^{\alpha(u-u_1)} 
\label{mtr}
\een
for $u_1 < u < u_0$, where $\alpha$ is a constant and the coefficients in \cref{mtr} were chosen so as to match the mirror position and velocity of \cref{acc} at $u=u_1$. In order that $v=0$ at $u=u_0$ we must have $p(u_0) = 0$, so we require
\ben
 \frac{1}{\alpha} \left( e^{\alpha(u_0-u_1)} - 1 \right) = \frac{1}{\kappa} \, .
\een
In order that the mirror be at rest when $v=0$, $u=u_0$, we must have $p'(u_0) = 1$, so we also require
\ben
e^{\alpha(u_0-u_1)} = e^{\kappa u_1} \, .
\een
These relations imply that
\ben
\alpha = \kappa \left( e^{\kappa u_1} - 1 \right) \approx \kappa e^{\kappa u_1} 
\een
where we have used $\kappa u_1 \gg 1$. We also have
\ben
u_0 \approx u_1 (1 + e^{-\kappa u_1}) \, .
\een
The mirror is then assumed to follow the inertial trajectory $p(u) = u-u_0$ for $u > u_0$. 

For the motion \cref{acc} followed by \cref{mtr}, and then followed by inertial motion, $p(u)$ and $p'(u)$ are continuous, but $p''(u)$ is discontinuous at $u=u_1$ and $u=u_0$. These discontinuities yield $\delta$-function contributions to $T_{uu}$, \cref{mse}. Taking these contributions into account, we have for $u_1 \leq u \leq u_0$, 
\ben
T_{uu} = \frac{1}{48 \pi} \alpha^2 - \frac{1}{24 \pi} (\alpha + \kappa) \delta(u-u_1) + \frac{1}{24 \pi} \alpha \delta(u-u_0)
\een
Thus, the total energy radiated in the ``burst'' between $u_1$ and $u_0$ is
\ben
E_B = \int_{u_1}^{u_0} T_{uu} du=  \frac{1}{48 \pi} \alpha^2 (u_0-u_1) - \frac{1}{24 \pi} \kappa \approx \frac{1}{48 \pi} \kappa^2 u_1 e^{\kappa u_1}
\label{ebrest}
\een
Thus, we have $E_B/E_H \approx \exp(\kappa u_1) \gg 1$. This is in accord with the estimate that would be obtained from our particle analysis 
(see \cref{paren}).

On the other hand, we can return the mirror to inertial motion by simply setting the acceleration of the mirror to zero at $u=u_1$, without changing its velocity, i.e., by having the mirror follow the trajectory \cref{mna}
for $u > u_1$. 
In that case, the only contribution to the energy flux for $u \geq u_1$ arises from the discontinuity in $p''$ at $u=u_1$, which yields
\ben
T_{uu} = - \frac{\kappa}{24 \pi} \delta(u - u_1)
\een
Consequently, the magnitude of the resulting (negative!) integrated energy flux is $\sim \kappa$, which is far less than the total Hawking energy flux \cref{thf}.

In summary, the purification of the Hawking radiation in the moving mirror spacetime of \cref{mmspacetime} always requires the emission of at least as many late time inertial particles as the total number of Hawking particles emitted. In the case where the mirror returns to rest by $v=0$, this requires a late time burst of energy far larger than the total energy contained in the Hawking radiation. However, in the case where the mirror transitions to inertial motion without changing its velocity, the purification of the Hawking radiation can occur with negligible energy cost.

\section{Vacuum Entanglement in Black Hole Evaporation}
\label{bh}

As discussed in the Introduction, if the semiclassical description is valid until the final (Planck scale) stage of black hole evaporation and if information is not lost in the process of black hole formation and evaporation, then the information stored within the black hole during the evaporation process must emerge at the final Planckian stage. The moving mirror example that we have analyzed in the previous sections suggests an interesting possibility as to how this might occur.

\begin{figure}[ht]
\centering
\begin{tikzpicture}
\draw (-2,-4)  -- (-2, 10.8);
%\draw (-2,-3)  -- (5, 4);
\draw (-2,-3)  -- (2, 1);
\draw [dashed] (2,1)  -- (5, 4);
\draw (12,-3)  -- (-2, 11);
\draw [very thick] (-2,1)  -- (2, 1);
%\draw (2,1)  -- (2, 4);
\filldraw [black] (2, 1) circle (2pt);
%\draw (-2,-1) .. controls (.5,-.8) .. (5,-1);
%\draw (2,2.5) .. controls (3,2.7) .. (5,2.5);
\node at (-2.7,0) {\textit{$r=0$}}; 
\node at (3.1,.8) {\textit{$r \sim l_P$}};
%\node at (1.3,3) {\textit{$r=0$}};
\node at (0,1.3) {\textit{high curvature}};
\node at (1.5,-1) {\textit{horizon}};
%\node at (4.5, 2) {\textit{light cone}};
\node at (1,4) {\text{Minkowski}};
%\node at (2.5,-.5) {\textit{$\Sigma_0$}};
%\node at (3.5,3) {\textit{$\Sigma_1$}};
\draw [blue] (-2,-3.4) -- (5.2, 3.8);
\draw [red] (-2,-2.6) -- (4.8, 4.2);
\draw [blue] (-2,-3.4) -- (-1.4, -4);
\draw [red] (-2,-2.6) -- (-.6, -4);
\node at (8.5,1.3) {\textit{$\mathscr I^+$}};
\draw [->] (2.5,.8) -- (2.05,.95);
\draw [->] (.8,.-1) -- (.2,-.8);
%\draw [->] (3.7,2) -- (3.2, 2.2 );
\node at (-.8,0) {\text{black hole}};
\node at (5.5, 3.9) {\color{blue}\textit{h}};
\node at (5.1, 4.4) {\color{red}\textit{$\bar{f}_1$}};
\end{tikzpicture}
\caption{A spacetime diagram of a black hole that forms by gravitational collapse and evaporates with no information loss (see text).}
\label{bhe}
\end{figure}
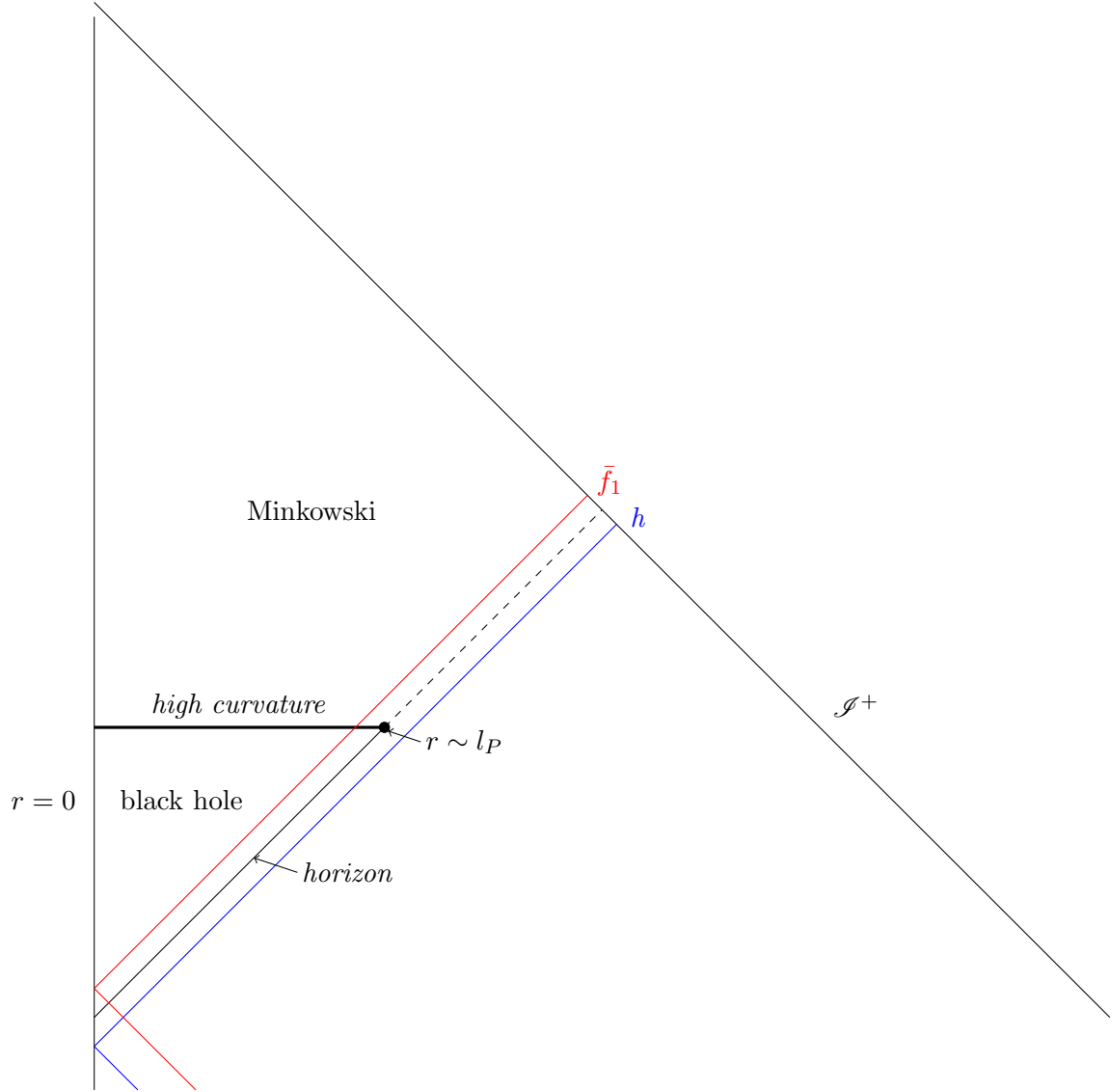

\Cref{bhe} is a conformal diagram of a black hole that forms by gravitational collapse and subsequently evaporates by emission of Hawking radiation, leaving behind empty, flat spacetime. (Note that there is considerable distortion of distances and times in \cref{bhe}.) \Cref{bhe} differs from the usual depiction of black hole evaporation in that the thick line we have labeled as ``high curvature'' normally would be shown as a singularity, and the final moment of evaporation normally would be depicted as a point rather than a sphere of order Planck radius $l_P$. (A new ``origin of coordinates'' would then normally be depicted as emerging from the evaporation event.) In such a usual spacetime depiction of black hole evaporation, information would necessarily be lost into the singularity. In order to avoid information loss without modifying the semiclassical description until Planck scale curvatures are reached, we must replace the singularity by a high curvature regime through which the quantum fields---or whatever describes matter and gravity in this regime---can propagate, as we have done in \cref{bhe}.

The spacetime path of a Hawking mode $h(u)$ of inertial frequency $\omega$ is depicted in blue in \cref{bhe}. Just as in \cref{mmspacetime}, the backward in time propagation of $h(u)$ yields the mode $\tilde{h}(v)$ at $\mathscr I^-$ (not shown in \cref{bhe}), although now the change in inertial frequency from $h(u)$ to $\tilde{h} (v)$ is produced by the gravitational blueshift associated with the collapse rather than produced by a Doppler blueshift from reflection by a mirror. The reflection of $\tilde{h}(v)$ about $v=0$ at $\mathscr I^-$ (where $v=0$ is the advanced time at which the horizon forms) yields a ``partner mode,'' whose spacetime path is shown in red in \cref{bhe}. In a semiclassical analysis, the Hawking particle will be entangled with this partner mode. As in the mirror case, the Hawking radiation is in a mixed, thermal state. If the partner mode were to propagate into a singularity, then the final state of the quantum field would be mixed, i.e., information would be lost. However, if, instead, the partner mode were to propagate through a high curvature regime and out to infinity as illustrated in \cref{bhe}, then information loss could be avoided.

We do not know the physics that would apply in the high curvature regime of \cref{bhe}, so we do not know the fate of the partner mode if/when it propagates through the high curvature region. But an interesting possibility is that---just as in the mirror spacetime of \cref{mmspacetime}---it becomes a Milne mode of Milne frequency $\omega$ in the final Minkowski region. (Note that the future of the evaporation ``event'' at $r \sim l_P$ depicted in \cref{bhe} is essentially a future light cone of Minkowski spacetime---with a Planck radius sphere instead of a point at the vertex of the cone.) In that case, in analogy with \cref{hm}, for any Hawking mode $h(u)$, the final ``out'' state would be of the form
\begin{equation}
\Psi =  \left(\sum_n e^{-n\pi \omega/\kappa} |n\rangle_{h}  |n\rangle_{f_1} \right) \otimes \Psi'
\label{hm2}
\end{equation}
where $f_1$ denotes the Milne particle mode entangled with the Hawking mode $h$. {\em In other words, if \cref{hm2} holds, then the information stored within the black hole during its evaporation would emerge at the end of the evaporation process as vacuum fluctuations in the Minkowski region that are entangled with the Hawking radiation.}

In our review \cite{uwrev}, Unruh and I considered this possibility to be a potentially viable way of restoring information in black hole evaporation. However, the analysis of the particle and energy cost of vacuum entanglement in the moving mirror spacetime with final state of the form \cref{hm} can be taken over directly to the black hole spacetime with final state of the form \cref{hm2}. By the same analysis as given in \cref{pecost}, the final state \cref{hm2} must contain as many late-time inertial particles as the total number of Hawking particles emitted during the evaporation process. In the mirror case, we could make the total energy associated with these late time particles very small compared with the energy emitted in Hawking radiation by turning off the acceleration of the mirror at the end of the Hawking process without changing its velocity, so that the late time particles have extremely low inertial frequencies. However, there is no plausible analog of this in the black hole case. By causality, the Milne modes $f_1$ depicted in \cref{bhe} must emerge from a Planck scale region of the final Minkowski portion of the spacetime. The inertial frequencies of these modes therefore must be of essentially Planck scale, and the corresponding inertial particles must be of essentially Planck energy. Unlike the moving mirror case, there is no ``external agent'' who can supply this energy. Thus, the final state \cref{hm2} is not energetically possible.

In summary, the purification of Hawking radiation via entanglement with vacuum fluctuations in the final Minkowski region provides an interesting possibility for avoiding information loss. However, our analysis shows that---just as in other previously considered scenarios where the information emerges in a ``final burst''---it requires the emission of as many Planck scale inertial particles as Hawking particles and it is not energetically possible.

\bigskip

\noindent
{\bf Acknowledgements} I wish to thank Bill Unruh and Stefan Hollands for many extremely valuable discussions during the course of this work. This research was supported in part by NSF grants PHY 15-05124 and PHY18-04216 to the University of Chicago.

\bibliographystyle{JHEP}

\end{document}